\documentclass{lncs}
\usepackage{hyperref}
\usepackage[none]{hyphenat}
\usepackage{theorem}
\usepackage{csquotes}

\begin{document}
\title{Assignment of Different-Sized Inputs in MapReduce\thanks{{This is a version that appeared as a brief announcement in International Symposium on Distributed Computing (DISC), 2014. More details appear in the technical report 14-05, Department of Computer Science, Ben-Gurion University of the Negev, Israel, 2014. This work was partially supported by the project Handling Uncertainty in Data Intensive Applications, co-financed by the European Union (European Social Fund) and Greek national funds, through the Operational Program \enquote{Education and Lifelong Learning,} under the program THALES, the Rita Altura Trust Chair in Computer Sciences, Lynne and William Frankel Center for Computer Sciences, Israel Science Foundation (grant 428/11), the Israeli Internet Association, and the Ministry of Science and Technology, Infrastructure Research in the Field of Advanced Computing and Cyber Security.}}}
\titlerunning{Brief Announcement: Assignment of Different-Sized Inputs in MapReduce}
\toctitle{Assignment of Different-Sized Inputs in MapReduce}
\author{Foto Afrati~\inst{1} \and Shlomi Dolev~\inst{2} \and Ephraim Korach~\inst{2} \and Shantanu Sharma~\inst{2} \and Jeffrey D. Ullman~\inst{3}}
\institute{
National Technical University of Athens, Greece.
\and Ben-Gurion University of the Negev, Israel.
\and Stanford University, USA.}
\maketitle
\date{}
\thispagestyle{empty}


\noindent \textbf{Reducer Capacity.} An important parameter to be considered in MapReduce algorithms is the \enquote{reducer capacity.} A \emph{reducer} is an application of the reduce function to a single $key$ and its associated list of $value$s. The \emph{reducer capacity} is an upper bound on the sum of the sizes of the $value$s that are assigned to the reducer. For example, we may choose the reducer capacity to be the size of the main memory of the processors on which the reducers run. We assume that all the reducers have an identical capacity, denoted by $q$.

\medskip
\noindent \textbf{Motivation and Examples.} We demonstrate a new aspect of the reducer capacity in the scope of several special cases. One useful special case is where an output depends on \emph{exactly} two inputs. We present two examples where each output depends on exactly two inputs and define two problems that are based on these examples.

\smallskip \noindent \textit{Similarity-join.} Similarity-join is used to find the similarity between any two inputs, \textit{e}.\textit{g}., Web pages or documents. A set of $m$ inputs (\textit{e}.\textit{g}., Web pages) $\mathit{WP}= \{wp_1, wp_2, \ldots, wp_m\}$, a similarity function $sim(x, y)$, and a similarity threshold $t$ are given, and each pair of inputs $\langle wp_x, wp_y\rangle$ corresponds to one output such that $sim(wp_x, wp_y) \geq t$. It is necessary to compare all pairs of inputs when the similarity measure is sufficiently complex that shortcuts like locality-sensitive hashing are not available. Therefore, it is mandatory to compare every two inputs (Web pages) of the given input set ($\mathit{WP}$).

\medskip \noindent \textit{Skew join of two relations $X(A,B)$ and $Y(B,C)$.} \textnormal{The join of relations $X(A,B)$ and $Y(B,C)$, where the joining attribute is $B$, provides the output tuples $\langle a, b, c\rangle$, where $(a,b)$ is in $X$ and $(b,c)$ is in $Y$. One or both of the relations $X$ and $Y$ may have a large number of tuples with the same $B$-value.}
\textnormal{A value of the joining attribute $B$ that occurs many times is known as a {\em heavy hitter}. In skew join of $X(A,B)$ and $Y(B,C)$, all the tuples of both the relations with the same heavy hitter should appear together to provide the output tuples.}

\smallskip
\noindent \textbf{Problem Statement.}
We define two problems where exactly two inputs are required for computing an output, as follows: (\textit{i}) \textit{All-to-All problem}. In the \emph{all-to-all} (\textit{A2A}) problem, a set of inputs is given, and each pair of inputs corresponds to one output. Computing common friends on a social networking site and similarity join are examples. (\textit{ii}) \textit{X-to-Y problem}. In the \emph{X-to-Y} (\textit{X2Y}) problem, two disjoint sets $X$ and $Y$ are given, and each pair of elements $\langle x_i, y_j\rangle$, where $x_i \in X, y_j \in Y, \forall i, j$, of the sets $X$ and $Y$ corresponds to one output. Skew join and outer product or tensor product are examples.

%

The \textit{communication cost}, \textit{i}.\textit{e}., the total amount of data transmitted from the map phase to the reduce phase, is a significant factor in the performance of a MapReduce algorithm. The communication cost comes with tradeoff in the degree of parallelism however. Higher parallelism requires more reducers (hence, of smaller reducer capacity), and hence a larger communication cost (because the copies of the given inputs are required to be assigned to more reducers). A substantial level of parallelism can be achieved with fewer reducers, and hence, yield a smaller communication cost. Thus, we focus on minimizing the total number of reducers, for a given reducer capacity $q$. A smaller number of reducers results in a smaller communication cost.

\medskip \noindent \textbf{Tradeoffs.} The following tradeoffs appear in MapReduce algorithms and in particular in our setting: (\textit{i}) a tradeoff between the reducer capacity and the total number of reducers, (\textit{ii}) a tradeoff between the reducer capacity and parallelism, and (\textit{iii}) a tradeoff between the reducer capacity and the communication cost.

\medskip \noindent \textbf{Mapping Schema.} A mapping schema is an assignment of the set of inputs to some given reducers under the following two constraints: (\textit{i}) a reducer is assigned inputs whose sum of the sizes is less than or equal to the reducer capacity, and (\textit{ii}) for each output, we must assign the corresponding inputs to at least one reducer in common. The following two problems are proved to be NP-compete:

\medskip \noindent \textbf{The \emph{A2A Mapping Schema Problem.}} An instance of the \emph{A2A mapping schema problem} consists of a set of $m$ inputs whose input size set is $W=\{w_1, w_2, \ldots, w_m\}$ and a set of $z$ reducers of capacity $q$. A solution to the \emph{A2A mapping schema problem} assigns every pair of inputs to at least one reducer in common, without exceeding $q$ at any reducer.

\medskip \noindent \textbf{The \emph{X2Y Mapping Schema Problem.}} An instance of the \emph{X2Y mapping schema problem} consists of two disjoint sets $X$ and $Y$ and a set of $z$ reducers of capacity $q$. The inputs of the set $X$ are of sizes $w_1, w_2, \ldots, w_m$, and the inputs of the set $Y$ are of sizes $w_1^{\prime}, w_2^{\prime}, \ldots, w_n^{\prime}$. A solution to the \emph{X2Y mapping schema problem} assigns every two inputs, the first from one set, $X$, and the second from the other set, $Y$, to at least one reducer in common, without exceeding $q$ at any reducer.

\end{document}